# Collaborative Driving: Learning-Aided Joint Topology Formulation and Beamforming

Yao Zhang, Northwestern Polytechnical University, Changle Li, Tom H. Luan, XidianUniversity, Chau Yuen, Singapore University of Technology and Design, Yuchuan Fu, Xidian University

Currently, autonomous vehicles are able to drive more naturally based on the driving policies learned from millions of driving miles in real environments. However, to further improve the automation level of vehicles is a challenging task, especially in the case of multi-vehicle cooperation. In recent heated discussions of 6G, millimeter-wave (mmWave) and terahertz (THz) bands are deemed to play important roles in new radio communication architectures and algorithms. To enable reliable autonomous driving in 6G, in this paper, we envision collaborative autonomous driving, a new framework that jointly controls driving topology and formulate vehicular networks in the mmWave/THz bands. As a swarm intelligence system, the collaborative driving scheme goes beyond existing autonomous driving patterns based on single-vehicle intelligence in terms of safety and efficiency. With efficient data sharing, the proposed framework is able to achieve cooperative sensing and load balancing so that improve sensing efficiency with saved computational resources. To deal with the new challenges in the collaborative driving framework, we further illustrate two promising approaches for mmWave/THz-based vehicle-to-vehicle (V2V) communications. Finally, we discuss several potential open research problems for the proposed collaborative driving scheme.

## Introduction

The growing popularity of autonomous vehicle is triggering a revolution of transportation systems according to a report from Transportation Research Board (TRB) that predicts autonomous vehicles may account for 90% of vehicle sales and 65% of vehicle travel by 2050 [1]. By replacing human drivers with AI-empowered computers, autonomous driving is moving vehicles towards intelligence, safety and efficiency. In this case, multi-vehicle cooperation will play a vital role as driving decisions of vehicles should be determined in a collaborative way by sharing knowledge and learned policies.

Autonomous driving is expected to reshape driving experiences in two stages. Specifically, in the short term, autonomous or partially autonomous vehicles will free drivers from boring and tiring driving behaviors. New vehicular applications, including in-vehicle voice assistant, AR-enabled heads-up displays, and AI-based parking, etc., will become the norm in vehicles. In the longer term, a blueprint of the future transportation system is envisioned by Daimler and Bosch [2]. In this blueprint, autonomous vehicles will drop off and pick up users, plan driving route, and choose car park in a fully automated way. As such, the functionality of vehicles can be extended, such as mobile offices or even bedrooms, in order to improve the productivity and travelling experience of users. From recent trend of autonomous vehicles, two features play important roles, i.e., big data and intelligence:

- **Big Data**: It is estimated that more than 200 sensors will be equipped in a future autonomous vehicle, which requires sensor bandwidth of 3 Gb/s to 40 Gb/s [3]. A similar prediction from Intel shows that about 4 TB data will be produced from on-board sensors in an autonomous vehicle [4]. Some of these sensors generate sensory data in the form of streaming, which significantly challenges the on-board computing system.
- **Intelligence:** With recent advances of artificial intelligence (AI) and data analytics technologies, vehicles are becoming intelligent. For example, the advanced driver assistance system (ADAS), which assists drivers to park, change driving lane, navigate, etc., is an early implementation of low-level automation In recent years, higher-level automation of vehicles becomes the main bottleneck in leading automotive manufacturers worldwide. In particular, the first autonomous vehicle on sale with Level 3 self-driving capability is Audi 2018 announced in 2018. Other automakers, such as GM, Ford, and Toyota, put their efforts to speed R&D for autonomous vehicles towards level-3 or level-4 automation [5] but a few of them have achieved the mass production.

To promote the automation of vehicles, it is vital to guarantee the sensing accuracy and efficiency when processing large volume of sensory data. Unreliable sensing systems are fragile according to many fatal autonomous driving crashes. For example, [6] reported six serious traffic crashes occurred by Google Self-Driving Car, Lexux SUV, Tesla Model S and Model X, Uber Self-Driving Volvo. A main reason in these crashes is due to sensing blind area and unreliable processing of sensory data. Multi-vehicle cooperation has been demonstrated as a promising solution to enable reliable driving of vehicles, e.g., cooperative adaptive cruise control (CACC) and platooning/clustering-based driving [7]. However, a new driving scheme of multi-vehicle cooperation is necessary to address the challenges of autonomous vehicles in vulnerable sensing and unreliable driving. Targeting on the higher-level automation of vehicles, this paper aims to design a collaborative driving architecture by investigating two problems:

Q1: *How to use multi-vehicle cooperation to extend sensing abilities of collaborative vehicles?*

Q2: *How to enable reliable and effective data sharing among collaborative vehicles?*

For Q1, we propose a new collaborative driving framework by extending platooning-based driving whose effectiveness has been proved from both academic community [7] and industry (e.g., Volvo[1], Peloton[2]). Specifically, in the proposed collaborative driving framework, the driving topology of a group of vehicles is formulated and maintained in order to drive the group vehicles in a uniform way. To improve sensing efficiency, we adopt centralized decision-making (in group leader) and decentralized execution of sensing tasks (in group members). In each driving group, the group leader makes group-level driving decisions for the group, including path planning, inter-group interactions, allocation of sensing tasks, etc., while the group members undertake single-vehicle tasks, including area-level decision making, intra-group cooperation, etc. The sensory data and preprocessing results of group members will be collected by the group leader to make driving decisions. As such, we divide decision-making tasks and sensing tasks to group leader and group members, respectively.

Technologies supporting autonomous driving can be divided into two types: computation-oriented and communication-oriented. The first one aims to promote driving accuracy and reliability, such as object detection, route planning, route planning, etc [8]. The second one aims to achieve gigabit-per-second vehicular communications, which however is difficult for existing vehicular communication protocols, such as dedicated short range communication (DSRC), long-term evolution-based vehicular communication (LTE-V), and 5G-V2X [9][10].

In recent heated discussions of 6G, wireless communication is moving towards mmWave/THz bands in order to support diversified AI-supported applications with high requirements on data transmissions and processing. On one hand, mmWave/THz-enabled communication improves the efficiency of short-range wireless transmissions.

On the other hand, it also makes the communication architectures and algorithms substantially different from that of traditional communication systems. Vehicular communications in mmWave/THz bands are more challenging because of harsh propagation conditions and blockage-rich environment.

We design two promising approaches to solve the challenges above in order to ensure that the proposed collaborative driving scheme enjoy "wire-like" performance in mmWave/THz bands. Firstly, to deal with the misalignment problem, an efficient way is to align mmWave/THz beams with the assistance of out-of-band WiFi information by measuring the channel response of probing packets and then providing the hint of dominatingpaths. As such, we develop a DSRC-assisted beam alignment method to improve the efficiency of beam alignment of V2V links. Secondly, graph neural networks (GNNs) are able to achieve excellent performance in learning features from non-Euclidean data. We thus develop GNN-based beamforming scheme, which uses GNN to learn important features of V2V links and predict best beamformer.

## A TYPICAL APPLICATION SCENARIO OF COLLABORATIVE DRIVING.

In this section, we provide an autonomous driving scenario as an example, i.e., public traffic service systems, to show the potential benefits of collaborative driving.

In the near future, fully autonomous vehicles will benefit firstly public traffic service systems, such as Uber[3], and DiDi[4]. This is because that the common goal of these systems is to maximize service efficiency with no sacrifice of data privacy, which is easily acceptable for users. As such, we define a service-oriented three-layer collaborative driving architecture, as shown in Figure 1. Three key layers are introduced below.

- **User Layer:** User layer provides users with the interface to interact with autonomous vehicles. To relax users in public traffic service systems, all those users need to do is to input instructions including source, destination, and other requirements of entertainment before payment. By self-decision making, autonomous vehicles execute all the procedures such as control, route planning, pick up, and drop off automatically. The specification of user layer is to enable users enjoying the convenient operation by regarding the service process of autonomous vehicles as a black box.

- **Decision Layer:** The group-level decisions are made in decision layer. Two processes, i.e., the centralized decision making of group-level driving tasks and decentralized area-level sensing tasks, are included in decision layer. That is, the group leader in a group makes driving decisions (e.g., route planning) for the group by processing the collected sensory data from group members while group members undertake area-level sensing tasks allocated from the group leader. Therefore, by deploying efficient multi-vehicle perception algorithms and task allocation policies, decision layer will make driving decisions with high computational and communication efficiency.

---

[1] https://www.volvogroup.com

[2] https://peloton-tech.com

[3] https://www.uber.com/

[4] http://www.didichuxing.com/en/

• **Control Layer:** The mobility control tasks of vehicles are managed in control layer. Specifically, a topology control strategy is developed in each group leader. Based on the control laws generated from the strategy, each vehicle decides their acceleration/deceleration actions and maintains proper inter-vehicle distance. Besides, the strategy updating is based on the driving decisions from decision layer. The management policies for group members, e.g., vehicle joining, are also included in the control level. Therefore, by adjusting the driving actions of each vehicle, control layer is important to stabilize group driving.

## FORMATION OF COLLABORATIVE DRIVING

In this section, focusing on the decision layer and control layer of the three-layer architecture presented earlier, we illustrate the principle of the new collaborative driving framework.

### Preliminary

In collaborative driving, sensing tasks can be distributed in collaborative vehicles, where each vehicle has a specific sensing range and the preprocessing results of sensory data are finally combined in the leader vehicle (group leader). In this way, the environmental sensing of vehicles in a group is distributed to a part of vehicleswhile the common driving decisions are made by the group leader, accordingly forming a hierarchal collaborative driving with cooperative sensing, as shown in Figure 2.

In what follows, after introducing the hierarchal cooperative sensing in the proposed collaborative driving scheme, the subsection of data processing is to describe the data types in collaborative driving. It should be noted that the collaborative driving must be supported by stable driving topology. Hence, topology formation that ensures the stable headway distance among vehicles is exploited for fully autonomous vehicles. In the final part of this section, the benefits of collaborative driving are extracted.

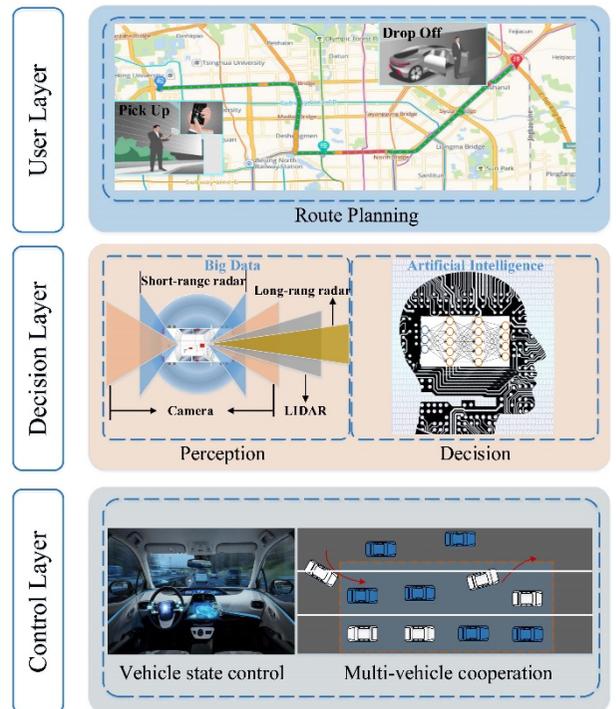

**Figure 1** Service-Oriented Three-Layer Collaborative Autonomous Driving Structure.

### Cooperative Sensing

In Figure 2, the group leader is denoted as the middle one (white) who makes group-level decisions, allocates sensing tasks and manages group members, i.e., periphery vehicles (green). The sum of $\theta_1$ to $\theta_2$ in Figure 2 is approximately equal to 360 degrees, which means that the periphery vehicles contribute an omni-directional sensing. Each periphery vehicle makes area-level decisions, i.e., preprocessing sensor data. At the same time, the group leader collects the preprocessed results of sensor data from group members and makes uniform driving decisions for the group. In this case, there is no need for the group leader to undertake sensing tasks if a full sensing view was covered by the periphery vehicles. In this pattern, the redundant sensor data of the group will be reduced significantly.

### Data Processing

We list four types of data in the proposed collaborative driving scheme and introduce their roles as following.

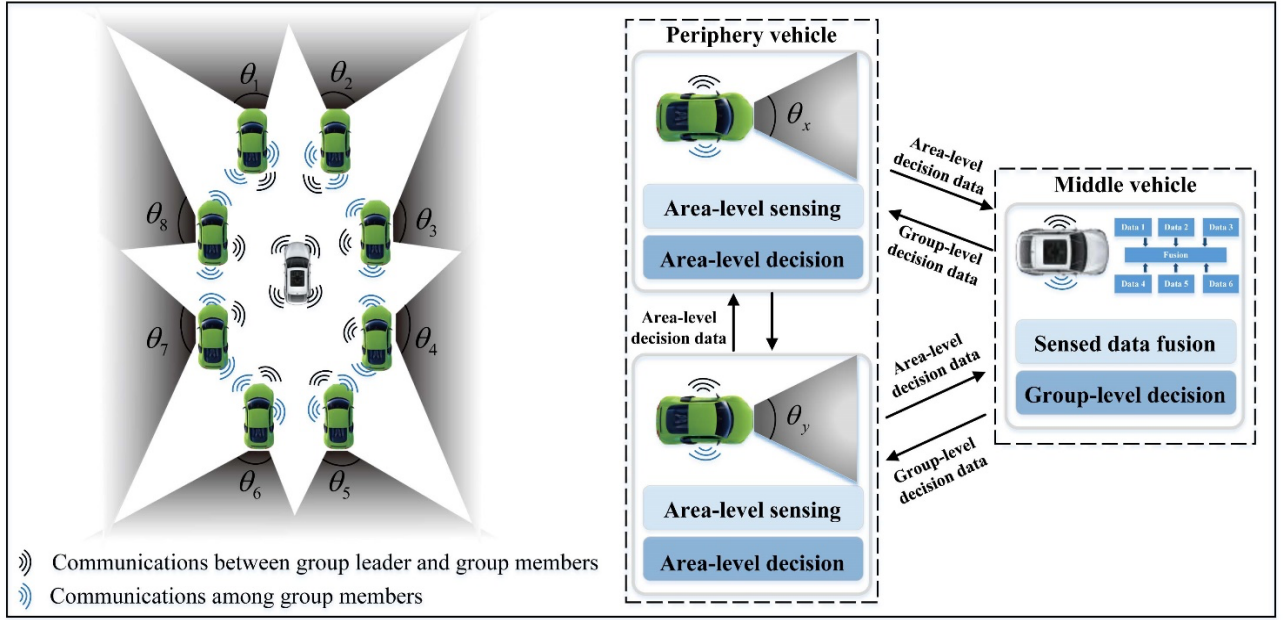

*Figure 2* Hierarchal Architecture of Collaborative Driving

- **Sensing data:** The area-level sensing data are first obtained and processed in periphery vehicles, and then transmitted to the group leader. The group leader extracts important features from the received full-view sensing results and then makes driving decisions for the group. This process consumes both the communication and computing resources.
- **Decision data:** Recall that the decision-making process includes two scales, i.e., area-level decisions and group-level decisions. In this case, the group-level decision data are delivered by group leader in a broadcast/multicast way to inform group members. The area-level decision data are transmitted from each group member to other group members or the group leader, in order to aggregate the independent sensing results from each group member.
- **Request data:** Request data are sent from group members or free vehicles, in order to request the admission of entering or leaving. Free vehicles refer to the vehicles that have no affiliation with any groups and are looking for a proper group to enter.
- **Control data:** The group leader manages the group through control data. In other words, control data is used to respond the entering and leaving requests of free vehicles and periphery vehicles. For topology control, the information related to road environments and driving directions is also carried by control data.

The implementation of collaborative driving is supported by several technologies, including data analytics, resource management, load balance, task allocation. Most of these technologies rely on underlying communication technologies. Therefore, in this paper, we explore the directional mmWave/THz communications in vehicular environments and leave others as open issues.

## Topology Control

This part illustrates the basic procedure of driving topology control as well as its requirements. Topology control is to stabilize a group of vehicles with a common driving goal, which is challenging in terms of headway management, topology maintainance in the processes of entering and leaving. We take the entering process an example and present the stepped topology control procedure as follows:

---

**Algorithm 1** Topology Formation Procedure

**Input:** Current topology statues, including number of group members, driving actions;
**Output:** Forming a new driving topology for group members.
1: A free vehicle broadcasts requests to enter driving group;
2: If there is no response, the free vehicle starts to develop a driving group and acts as a group leader;
3: After receiving request information, the group leader updates the topology by determining positions of the free vehicle and respond it;
4: The free vehicle merges into the group once it receives the admission information, and then acts as a new periphery vehicle;
5: The predecessor periphery vehicle is converted to an idle one if necessary, which acts as the forward node between the new periphery vehicle and group leader.
6: The new driving topology is formed and the group leader broadcasts the topology updating to all members.

---

Be similar with the entering process above, the leaving process is to stabilize the driving topology when group members leave the group. Specifically, if the leaving vehicle belongs to periphery vehicles, the group leader modifies the idle vehicle attached with the leaving vehicle to a new periphery vehicle. If there is no attached idle vehicles, the sensing range of existing periphery vehicles will be altered to cover the lost sensing area. When the leaving vehicle is a group vehicle, the nearest vehicle will become a new group leader to stabilize the group.

*Benefits of Collaborative Driving*

The proposed collaborative driving scheme enables autonomous vehicles benefiting in four particular aspects, i.e., sensing efficiency, resource utilization, traffic efficiency, and deployment efficiency. This is because the multi-vehicle cooperation can achieve task partition and resource integration for environmental sensing.

- **High-Level Cooperation:** The cooperative sensing over the collaborative driving framework can avoid uneven sensing and diverse processing abilities in autonomous vehicles. This is because the sensing tasks allocated by the group leader are on-demand and area-partitioning. As a result, the computing and communication resources of group members will be utilized in a high-efficiency way.
- **Sensing Efficiency:** In existing autonomous driving systems, vehicles often drive independently and have no cooperation with each other. It is naturally for autonomous vehicles to generate redundant sensory data and sensing blind area due to blockage. With the topology control, the driving topology can be continuously optimized to minimize sensing redundancy with guarantee of sensing coverage. By designing partition and allocation policies of sensing tasks, the overall sensing efficiency could be improved.
- **Resource Utilization:** Autonomous driving should be supported by high-reliability communication technologies and high-performance computing capabilities, in order to meet the requirements in dealing with data stream collected from dynamic traffic environments. However, if not managed properly, resource waste will degrade the computing and communication efficiency. The topology control policy in collaborative driving can be optimized to achieve optimal V2V association that ensures minimum spectrum interference and maximal utilization of computing resources since vehicles usually have diverse computing capabilities.

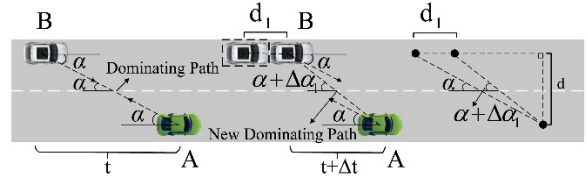

**Figure 3** Beam alignment for the azimuth angle of moving transmission pairs

- **Traffic Management:** Traffic efficiency and safety are two important goals of autonomous driving. Unfortunately, existing autonomous vehicles have little impact on traffic efficiency due to existence of semi-autonomous driving and human driving. The effectiveness of platooning has been proved in many practical scenarios [7]. Based upon platooning, the proposed collaborative driving has a potential to bring overwhelming advantages in traffic efficiency and safety.
- **Deployment Efficiency:** The AI based autonomous vehicles must be trained and tested for hundreds of millions of miles before deployment [11], in order to deal with unknown cases in all kinds of traffic scenarios. As such, the huge training costs limit the fast application and development of autonomous vehicles. With topology control, the collaborative driving framework is able to provide an underlying platform for the application of distributed learning algorithms, such as federated learning. Sharing both raw sensory data and inference knowledge in real time will speed the training process of neural network models and thus improve the performance on inference.

## ULTRA-WIDE BANDWIDTH FOR V2V COLLABORATIVE DRIVING.

In this section, we investigate the application of mmWave/THz-based V2V communications in collaborative driving by providing DSRC-assisted beam alignment method and then propose a GNN-based beamforming for V2V communications.

*DSRC-Assisted Beam Alignment*

In the radio communication systems of 6G, mmWave/THz will play a vital role, but also increase the complexity of system. Efficient beam alignment in such wide bandwidth is a challenging task because of link dynamics and possible blockage. Different from traditional beam alignment methods, such as the

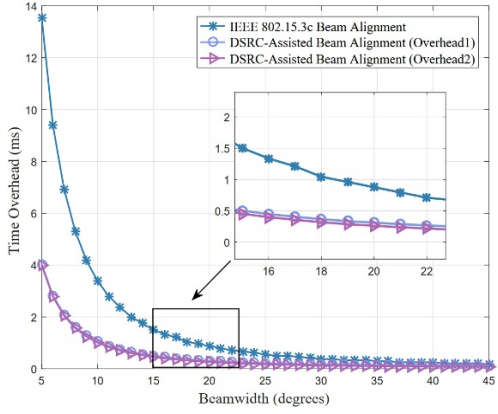

*Figure 4* Overhead of two beam alignment schemes

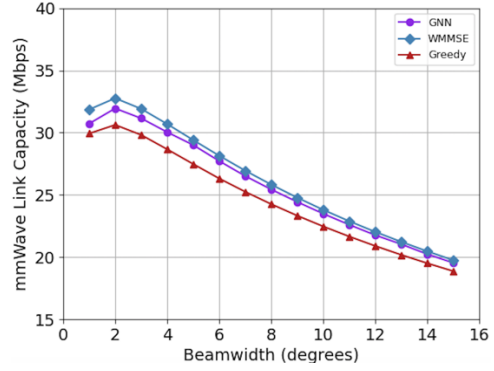

*Figure 5* mmWave link capacity with different beamforming angles searching algorithms in IEEE 802.15.3c, out-of-band alignment with extra GPS information [9], we design a DSRC-assisted beam alignment method for mmWave communication, in order to achieve fast and cost-effective beam alignment. To overcome the impact of vehicle mobility on beam alignment, we further develop a two-step convolution scheme by incorporating the movements of both communicating pairs based on [12]. Particularly, DSRC enables robust channel links but with lower data rate than mmWave. As such, our method uses DSRC channel information to guide the beam alignment and reduce extra pointing overhead.

Considering a typical scenario shown in Figure 4, where the mmWave link between communication pairs (the vehicle *A* and B) is tracked by their movements, the two-step beam alignment is shown as follows.

• **First-step convolution:** For vehicle *A* in Figure 3, the direction of propagation path is assumed to be shifted from ($\alpha$, $\beta$) to ($\alpha+\Delta\alpha_1$, $\beta+\Delta\beta_1$) due to the movement of vehicle *B*, where $\alpha$ and $\beta$ are the azimuth and elevation angles, respectively. The angular shift in the dominating propagation direction can be obtained by the angular profile of successive DSRC frames. Therefore, we can obtain the best beam of vehicle *A* by convolving the beam gain $G_{b1}(\alpha, \beta)$ and the channel gain $G_{c1}(\alpha+\Delta\alpha_1, \beta+\Delta\beta_1)$ in the dominating propagation path. Note that the angular shift $\Delta\alpha_1$ and $\Delta\beta_1$ are signed values. After that, the first-step convolution is completed. The beam alignment for azimuth angle considering the movement of vehicle *B* is shown in Figure 4. The alignment of elevation angle is similar but $\Delta\beta$ is typically zero if there is no slope on the road.

• **Second-step convolution:** The second step is to guide the angular shift of vehicle *B* based on the movement of vehicle *A*. After movement, the angular shift of the propagation path from vehicle *A* is ($\Delta\alpha_2$, $\Delta\beta_2$). The optimal beam of vehicle *B* thus can be obtained by convolving $G_{b2}(\alpha, \beta)$ and the channel gain $G_{c2}(\alpha+\Delta\alpha_2, \beta+\Delta\beta_2)$ based on the angular profile of successive DSRC frames. The second-step convolution is completed.

• **Final beam gain:** Finally, we can conclude that the angular shift of the dominating propagation path is ($\Delta\alpha_1+\Delta\alpha_2$, $\Delta\beta_1+\Delta\beta_2$), and the best beam is obtained by the convolution of $G_b(\alpha, \beta)$ and $G_c(\alpha+\Delta\alpha_1+\Delta\alpha_2, \beta+\Delta\beta_1+\Delta\beta_2)$. It should be noted again that the angular shift in $G_c$ is from DSRC frames.

The performance evaluation of the new beam alignment scheme is shown in Figure 4. Specifically, we assume that mmWave V2V communications can directly take place among all vehicles within each autonomous driving group and each time slot is composed of beam alignment slot and data transmission slot. Figure 4 presents the time overhead results of beam realignment, which denotes the beam realignment time when the beam misalignment occurs due to the movement of communication pairs. We compare two beam alignment schemes with varying beamwidths. We only focus on the beam-level alignment and assume the other two stages (device-to-device linking and sector-level searching) in IEEE 802.15.3c are previously established [13]. The *Overhead Scheme 1* in Figure 4 includes both the alignment time and DSRC overhead, while *Overhead Scheme 2* presents only the alignment time. The figure shows the performance gap of beam alignment overhead between DSRC-assisted alignment scheme (Overhead Scheme 2) and that of IEEE 802.15.3c is from 28.27% to 70.55%, corresponding to the beamwidth of 5dB and 45 dB. The time overhead decreases as the beamwidth increases because of the trade-off between alignment and throughput, i.e., narrow beamwidths can achieve high transmission rates but result in high beam-alignment overhead, and vice versa. The results of performance evaluation in Figure 4 illustrates the effectiveness of the proposed beam alignment method, which can be taken as a baseline in the future design of beam alignment solution for collaborative driving.

## GNN-Based Beamforming

To improve decision-making efficiency, we propose to adopt GNN to solve the complex beamforming problem in mmWave/THz-enabled V2V communications. This is because that GNN has an excellent performance in learning features from non-Euclidean data.

The beamforming problem in vehicular environments is challenging because of the dynamic inter-vehicle connections. Traditional solutions are optimization-based methods [14]. Considering that the inter-link relationships in vehicular communications belong to non-Euclidean space, GNN is expected to solve the beamforming problem. Specifically, Let $B$ denote the bandwidth of the transmitted signal, the goal of beamforming scheme is to maximize the link capacity, denoted as:

$$maximize_{P,A} \ Blog_2(1 + SNR),$$

where $P = \{p_1, p_2, ... p_N\}$ denotes the set of beamformers of V2V links and $p_1 = 1$ manes the $link_l$ is activated. $A = \{a_1, a_2, ... a_N\}$ denotes the set of link beamwidth, which is set to $\{1,2,…15\}$ degree in our experiments. To obtain signal-to-noise-ratio (SNR), we resort to the directional cone antenna model from [15] which assumes perfect antenna radiation pattern with signal energy concentrated uniformly in the generated beam. We then model the maximization problem above using graph representation. Briefly, after modeling the graph structure data as a graph $(V, E)$, GNNs are able to utilize the node features and the edge features $(E)$ to learn the representation of the vertices $V$. Based on the layer-wise structures, GNNs update the representation of each node by aggregating features from the hooked edges and the corresponding vertices. We use the message passing graph neural networks (a favorable family of GNNs) to accomplish the above updating process, described as:

$$x_i^{(k)} = \gamma^{(k)}\left(x_i^{k-1}, \Delta_{j \in N(i)} \phi^k\left(x_i^{k-1}, x_j^{k-1}, e_{j,i}\right)\right),$$

where $\Delta$ denotes a differentiable, permutation invariant function, e.g., sum, mean, or max, $\gamma$ and $\phi$ denote differentiable functions such as MLPs (Multi-Layer Perception). In our experiment, we set $\Delta$ as MAX function and $\gamma$ and $\phi$ as MLPs, respectively. We take each V2V link as the vertex in the graph, i.e., $m$-th vertex denotes the $m$-th transmitter-receiver pair. The vertex feature $x_i^{(k)}$ denotes the channel state of the corresponding V2V link at the $k$-round updating while the edge feature $e_{j,i}$ denotes the interference state of two links.

In our experiments, we set a 3-layer GNN and take the negative summation of link capacity as loss function. Therefore, the GNN-based beamforming is an unsupervised learning process so that it has no need to set labels. We use the popular adam optimizer and set the learning rate to 0.001. We set $\gamma$ to a MLP with hidden units of $\{5,32,32\}$ while the MLP in analogy to $\phi$ has the hidden units of $\{35, 16, 1\}$. The experiment results are shown in Figure 5. Although our GNN-based scheme cannot outperform WMMSE in terms of link capacity, it has a significant improvement in terms of computation efficiency. In our experiments with 12 moving vehicles, GNN inference takes millisecond-level time while WMMSE [14] takes second-level time.

## OPEN PROBLEMS AND SUMMARY

### Open Problems:

- Communication protocol design for collaborative driving: Different from existing vehicular communication protocols, e.g., IEEE 801.11p or C-V2X, mmWave/THz-based V2V communication highly relies on topology control when build reliable beam alignment schemes. For example, the design of MAC layer should be sensitive to driving topology and mobilities of vehicles in order to cater for the topology control.
- Distributed learning algorithms to optimize both communication and computational resources: Distributed leaning is able to further alleviate the computational burden of group leader while protect the data privacy of group members. However, how to design distributed learning algorithms by considering mmWave/THz bands and sensing tasks is challenging.
- New autonomous driving related applications: As collaborative driving provides a basic framework for multi-vehicle cooperation, more useful applications could be deployed. For example, vehicular AR will be developed based on the distributed data processing, which can extend the view of each vehicle that may be obstructed by other vehicles.

### Summary

Although autonomous driving technologies have been studied from academia and industry with considerable effort in recent years, we witness those fatal crashes are still inevitable. This motivates us to promote driving efficiency and safety by considering multi-vehicle cooperation. Therefore, a new collaborative autonomous driving framework is proposed. The framework provides centralized decision-making for group-level driving while sensing tasks are undertook by collaborative vehicles in a swarm-intelligence way. We also provide two cost-effective methods to improve mmWave/THz-enabled V2V communications. The proposed collaborative driving scheme is expected to provide an insight in the design of multi-vehicle cooperation based autonomous driving.

## Acknowledgments

This work was supported by the National Key R&D Program of China (2019YFB1600100), National Natural Science Foundation of China (U1801266 and 62101401), the Youth Innovation Team of Shaanxi Universities, A*STAR under its RIE2020 Advanced Manufacturing and Engineering (AME) Industry Alignment Fund – Pre Positioning (IAF-PP) (Grant No. A19D6a0053). Any opinions, findings and conclusions or recommendations expressed in this material are those of the author(s) and do not reflect the views of A*STAR. The corresponding author is Changle Li.

**Yao Zhang** (yaozh.g@nwpu.edu.cn) *is currently a Post-Doctoral Researcher with Northwestern Polytechnical University, Xi'an, China. He received the Ph.D. degree in Telecommunication Engineering from Xidian University, Xi'an, China, in 2020. He was a Research Assistant and Post-Doctoral Fellow with The Hong Kong Polytechnic University in 2019 and 2021, respectively. His current research interests include edge intelligence, networked autonomous driving.*

**Changle Li** (clli@mail.xidian.edu.cn) *received the Ph.D. degree in communication and information system from Xidian University, China, in 2005. He conducted his postdoctoral research in Canada and the National Institute of information and Communications Technology, Japan, respectively. He had been a Visiting Scholar with the University of Technology Sydney and is currently a Professor with the State Key Laboratory of Integrated Services Networks, Xidian University. His research interests include intelligent transportation systems, vehicular networks, mobile ad hoc networks, and wireless sensor networks.*

**Tom H. Luan** (tom.luan@xidian.edu.cn) *received Ph.D. degree from the University of Waterloo, Ontario, Canada, in 2012. He is now a professor at the School of Cyber Engineering of Xidian University, Xi'an,China. His research mainly focuses on content distribution andmedia streaming in vehicular ad hoc networks and peer-topeer networking, as well as the protocol design and performance evaluation of wireless cloud computing and edge computing.*

**Chau Yuen** (yuenchau@sutd.edu.sg) *received the B.Eng. and Ph.D. degrees from Nanyang Technological University, Singapore, in 2000 and 2004, respectively. He was a Postdoctoral Fellow with Lucent Technologies Bell Labs, Murray Hill, NJ, USA, in 2005, and a Visiting Assistant Professor with The Hong Kong Polytechnic University, Hong Kong, in 2008. From 2006 to 2010, he was with the Institute for Infocomm Research (I2R), Singapore. Since 2010, he has been with the Singapore University of Technology and Design, Singapore.*

**Yuchuan Fu** (ycfu@xidian.edu.cn) *eceived the Ph.D. degree from the School of Telecommunications Engineering, Xidian University, Xi'an, Shaanxi, China, in 2020. She is currently a Lecturer with the State Key Laboratory of Integrated Services Networks (ISN), School of Telecommunication Engineering, Xidian University. From 2018 to 2019, she was a joint Ph.D. Student with Carleton University, Ottawa, ON, Canada. Her current research interests include algorithm design in vehicular networks and autonomous driving.*